\documentclass[aps,prb,twocolumn,groupedaddress]{revtex4-1}

\usepackage{graphicx}
\usepackage{dcolumn}
\usepackage{bm}
\usepackage{units}
\usepackage{color}
\usepackage{natbib}

\begin{document}


\title{Substrate-Controlled Magnetism: Fe Nanowires on Vicinal Cu Surfaces}

\author{D. Hashemi$^1$}
\email{hashemi@umich.edu}
\author{M. Waters$^1$}
\author{V. S. Stepanyuk$^3$}
\author{W. Hergert$^2$}
\author{J. Kieffer$^1$}
\affiliation{$^1$University of Michigan, Department of Materials Science and Engineering, Ann Arbor, MI 48109 USA\\
$^2$Institut f\"{u}r Physik, Martin-Luther-Universit\"{a}t, Halle-Wittenberg, Von Seckendorff Platz 1, D-06120 Halle, Germany\\
$^3$Max-Planck-Institut f\"{u}r Mikrostrukturphysik, Weinberg 2, D-06120 Halle, Germany.}


\date{\today}

\begin{abstract}
	
Here we present a novel approach to control magnetic interactions between
atomic-scale nanowires. Our \emph{ab initio} calculations demonstrate
the possibility to tune magnetic properties of Fe nanowires formed
on vicinal Cu surfaces. Both intrawire and interwire magnetic exchange
parameters are extracted from DFT calculations. This study suggests
that the effective interwire magnetic exchange parameters exhibit
Ruderman-Kittel-Kasuya-Yosida- like (RKKY) oscillations as a function
of Fe interwire separation. The choice of vicinal Cu  surface
offers possibilities for controlling the magnetic coupling.
Furthermore, an anisotropic Heisenberg model was used in Monte Carlo
simulations to examine the stability of these magnetic configurations
at finite temperature. The predicted critical temperatures
of the Fe nanowires on Cu(422) and Cu(533) surfaces are well-above
room temperature.

\end{abstract}

\maketitle
%
%
\section{Introduction}

The continued need for increasing the information storage content of high density magnetic recording devices requires the development of new nanostructured magnetic materials such as chains, one-dimensional (1D) periodic linear arrangements of atoms. Most of the experimental methods and potential industrial applications require a high packing density of these chains. 1D periodic linear chains have been investigated experimentally \cite{shen1,shen2,mo2,Cyrus,Gambardella} and theoretically.\cite{lds08,lsw03,tung,mbb07,mzk08,sh02,sh03,sh03a,sh04a,mo1}

Stepped surfaces are common templates for 1D nanostructures\cite{teg09} since they can take advantage of 1D symmetry provided by an array of parallel steps on a vicinal surface. Cu surfaces can be prepared with a large number of atom-high steps through a procedure known as step decoration. In this process, material is deposited on a stepped surface and subsequently nucleates along the edges of the steps with chains or nanostripes  growing on the lower terraces along ascending step edges. However, Shen \emph{et al.}\cite{shen1,shen2} demonstrated that Fe nanostripes grow on the upper terraces of  stepped Cu(111) surfaces.

\begin{figure}[th!]
\centering
\includegraphics[trim = 0cm 0.3cm 0.1cm 1.7cm, clip, scale=1.1, angle=0]{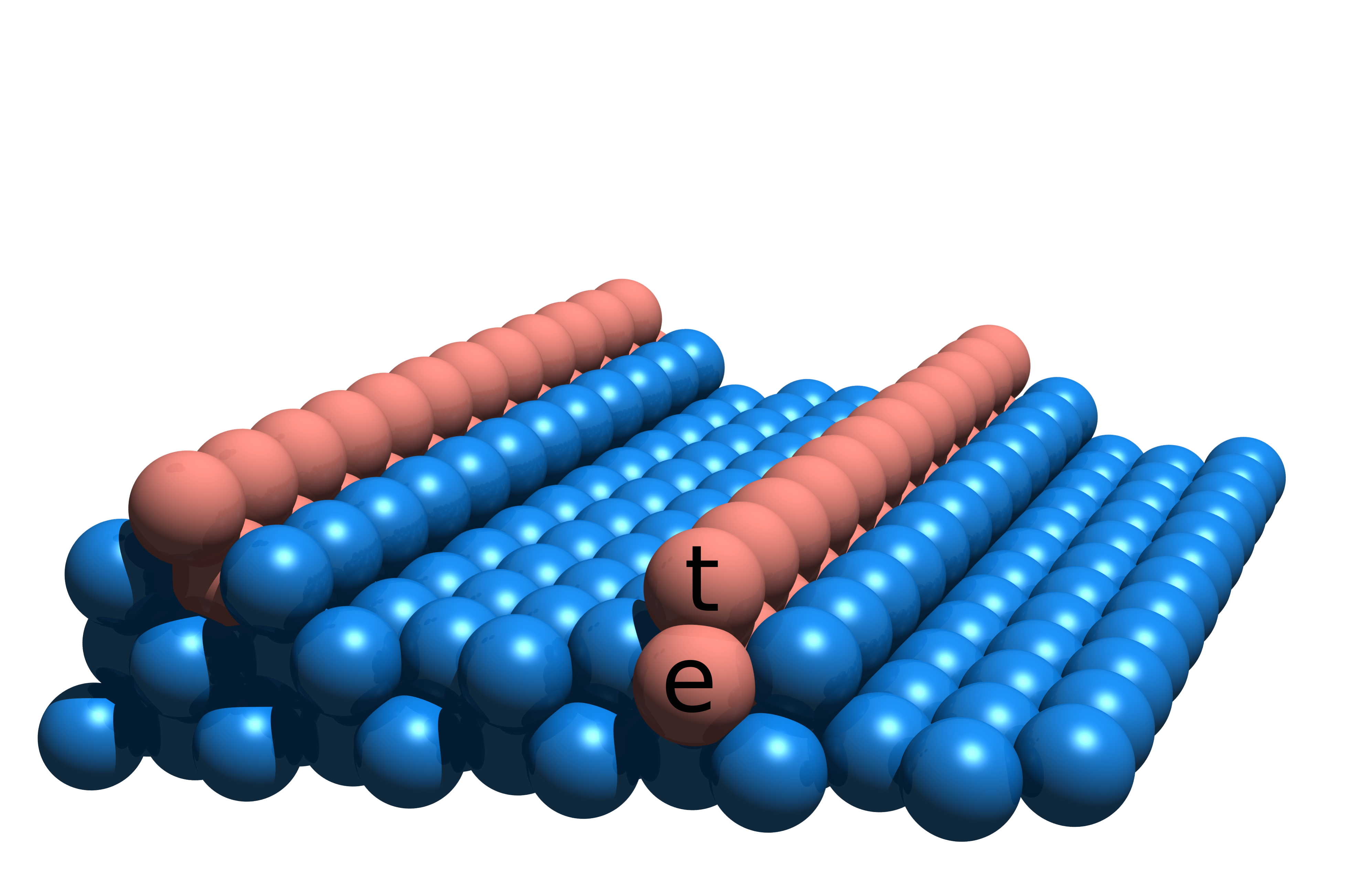}
\caption{ \label{fig:1} A one-atom wide Fe double chain (brown) is formed on vicinal
Cu(111) surface (blue). Fe chains at the top (t) and embedded (e) positions are also shown.}
\end{figure}

In an important study of the growth of linear Fe nanostructures on a stepped Cu(111) surface, Mo \emph{et al.}\cite{mo1} examined elementary diffusion and exchange processes  of Fe atoms on the surface by means of \emph{ab initio} calculations based on density functional theory (DFT). This study demonstrated the existence of a special two-stage kinetic pathway leading to the formation of Fe nanowires. In the first stage, Fe adatoms form a very stable 1D atom chain embedded in the Cu substrate behind a row of Cu atoms on the descending step. In the second stage, the embedded Fe chain acts as an attractor for subsequent Fe atoms deposited on the surface since Fe-Fe bonds are stronger than Fe-Cu bonds. This attraction assists in the formation of a secondary chain of Fe atoms on top of the original embedded Fe chain (cf. \ref{fig:1}) resulting in a very stable two atom-wide iron nanowire formed on the Cu surface. Total energy calculations revealed that the position of the Fe chain at the upper edge is energetically favorable to a Fe chain located at the step edge only if another row of Fe atoms is incorporated underneath the exposed row. \cite{mo1}

In a scanning tunnelling microscopy (STM) investigation aided by DFT calculations, Guo \textit{et al.} \cite{mo2} confirmed this growth process. A careful study of all atomic processes in the line of Ref.[\onlinecite{mo1}] has been used to perform kinetic Monte Carlo calculations. \cite{negulyaev2008} The simulations demonstrated the growth process as predicted by Mo \emph{et al.} and has been proven experimentally. \cite{mo2}

The interplay between dimensionality, local environment and magnetic properties has attracted special interest in such systems. In the following, a single linear periodic arrangement of atoms is referred to as a chain, while two parallel chains, either isolated or embedded in the Cu(111) surface, are called a wire.

The present investigation provides a systematic discussion of magnetic properties of 1D Fe nanostructures grown on a vicinal Cu(111) surface using the above mentioned template(cf. Fig. \ref{fig:1}). Detailed information on the real structure and magnetic states of such systems is given in Ref.[\onlinecite{hashemiprb,hashemijmmm}]. Ferromagnetic ordering is achieved for Fe wires deposited on this template. We present a systematic investigation of the magnetic couplings for Fe embedded in the Cu surface with terraces ranging from three to eight lattice constants wide. The analysis of the exchange coupling and of the magneto crystalline anisotropy allows us to set up a classical Heisenberg model to study finite temperature effects.

The outline of this paper is as follows. Sec. \ref{sec:Theoreticalmethod} is devoted to a brief description of the theoretical framework and setup we have used. Exchange parameters extracted from the DFT calculations are discussed in Sec. \ref{sec:Magneticexchangeinteractions}. The magnetic phase transition to the paramagnetic state and an adequate estimation of the critical temperature on the basis of numerical simulations is discussed in Sec. \ref{sec:MagnetismatFiniteTemperatures}. Finally, in Sec. \ref{sec:Conclusions} we summarize our main results and conclude.

%
%

\section{Theoretical method\label{sec:Theoreticalmethod}}

The calculations were performed within the framework of
spin-polarized density functional theory, using the Vienna \emph{ab initio}
simulation package (VASP) \cite{vasp1,vasp2}. The frozen-core
full-potential projector augmented-wave method (PAW) was used
\cite{paw1}, applying the generalized gradient approximation of
Perdew, Burke, and Ernzerhof (GGA-PBE) \cite{PhysRevLett.77.3865}.

Computational details and convergence checks are the same as those in our previous study.\cite{hashemiprb} with minor changes that are explained in Sec. \ref{sec:Resultsanddiscussion}.

A supercell containing twelve Cu layers, corresponding to between 72 and 192 Cu atoms for Cu(n+2,n,n), with n=2-7, was constructed to model the Cu(111) stepped surface. Conversely, the terraces range from three to eight lattice constants wide. The distance from one slab to its nearest image was equivalent to 13.5\AA. The number of k points was chosen according to the requirement that the number of atoms times the number of k points in the irreducible Brillouin zone.

\section{Results and discussion\label{sec:Resultsanddiscussion}}

\subsection{Real structure of embedded Fe wires \label{sec:RealstructureofembeddedFewires}}


We compared the relaxation of a Cu(111) surface with an embedded Fe chain with that of a clean Cu(111) surface. The extent of relaxation in the second subsurface layer is generally small. The relaxation in the first subsurface layer is larger but only significantly so at the step edge. The relaxation of the surface layer of Cu(111) is dominated by lateral and inward relaxations. Lateral relaxation is directed towards the center of the terrace, causing compression. The lateral relaxation in the middle of the terrace is small. In general the surface layer shows an inward relaxation, which is large at the step edge. Together with the outward relaxation for the first Cu atom of the terrace, the relaxations reduce the interatomic distances at the step edge. Significantly larger relaxation is observed when one row of Cu atoms is substituted by one row of Fe atoms behind the step edge.  From a structural point of view the Fe chain acts as a ”center of attraction”.  On clean Cu(111) surfaces, in the center of a terrace, practically no lateral shift can be seen, whereas a Cu atom at the same site will shift towards the embedded Fe chain. The Cu atoms at the step edge are also strongly attracted to the Fe chain. The inward relaxation of the Fe chain is much larger than the corresponding relaxation of a Cu atom at this site.  In summary, the Fe chain dramatically increases   the tendency  for compression near the step. The predominant structural reorganization is an inward relaxation of the Fe atoms relative to their ideal positions. For the Fe wire the inward relaxation of Fe atoms at the top and embedded positions are 22.5 \% and 5.9 \%, relative to the Cu lattice plane distance, respectively.

\subsection{Magnetic exchange interactions \label{sec:Magneticexchangeinteractions}}

The analysis of the exchange couplings and of the magneto crystalline anisotropy allows to set up a classical Heisenberg model to study finite temperature effects in Sec. \ref{sec:MagnetismatFiniteTemperatures}.

\begin{figure}[th!]
\centering
\includegraphics[trim = 0cm 2.8cm 0cm 8.4cm, clip, scale=0.17, angle=0]{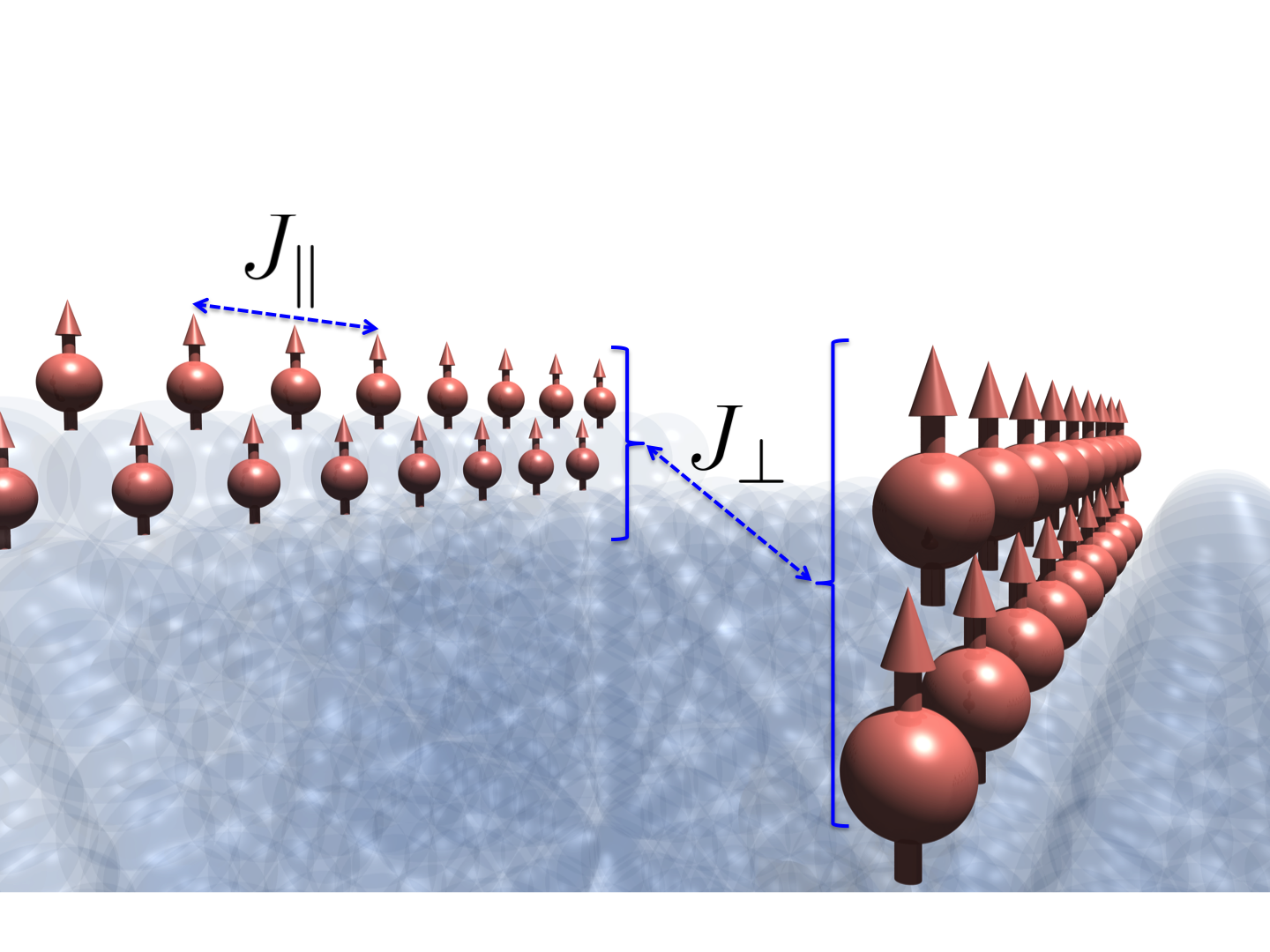}
\caption{ \label{fig:2} Schematic picture showing magnetic interactions of Fe wires. $J_{\parallel}$: Intrawire exchange coupling, and
$J_{\perp}$: Interwire exchange coupling.}
\end{figure}

For the Fe wires grown on vicinal Cu(111) surface, the absolute magnetic moments of Fe atoms at the top and embedded positions are 2.41 $\mu_B$ and 2.94 $\mu_B$, respectively.\cite{hashemiprb} There are two magnetic interactions between these moments: the intrawire ($J_{\parallel}$) and interwire ($J_{\perp}$) magnetic couplings (as shown in Fig.\ref{fig:2}) , both of which will be explored in this study.

%

The  Heisenberg  theory  of  magnetism maps magnetic interactions in a material onto localized  spin  moments. The resulting classical Hamiltonian,

\begin{equation}
H=-\sum_{i \neq j} J_{ij} \mathbf{e}_i \cdot \mathbf{e}_j - \sum_{i} K_{i} (\mathbf{e}_i \cdot \mathbf{e}_K)^2
\label{equ:aHeisenbergHamiltonian}
\end{equation}

contains the unit vectors $\mathbf{e}_{i(j)}$ of the magnetic moments, the  exchange  parameters $J_{ij}$, the magnetic anisotropy energy (MAE) $K_{i}$ (at site $i$), and the unit vector along the magnetization easy axis $\mathbf{e}_K$. Here $i$ and $j$ index the sites.

The MAE is 4.76 meV per site. \cite{hashemianisotropy} Using a constant value for MAE does not have any effect on the calculation of exchange couplings since our previous study showed that the MAE of Fe and the embedded Fe sublattices are equal. \cite{hashemianisotropy}

$J_{ij}$ can be calculated by making parallel and antiparallel alignment of the moments. Therefore,

\begin{equation}
 J_{ij} = \frac{H_{AF} - H_{FM}}{2}
\label{equ:aHeisenbergHamiltonian}
\end{equation}

where $H_{AF}$ and $H_{FM}$ are the DFT total energies calculated for antiparallel alignment of the moments and parallel alignment of the moments, respectively.

The interwire coupling constants are calculated in a  similar manner by exploiting supercells doubled in the direction perpendicular to the wires, and for parallel and antiparallel alignment
of the moments on the two wires on each side of the supercell.

\subsubsection{$J_{\parallel}$: Intrawire exchange coupling \label{sec:Intrawireexchangecoupling}}

In order to systematically study the intrawire exchange coupling of Fe wires, three systems were investigated; freestanding Fe chains, freestanding Fe wires, and embedded Fe wires. In freestanding Fe chains the atomic distances were constrained to the Cu bond length of the Cu(111) substrate in order to simulate a freestanding equivalent to a singular Fe chain in the Fe wire on the substrate. A freestanding wire was studied as an equivalent to the one embedded into the Cu(111) surface. All interatomic distances also correspond to the Cu bond length of the substrate in this case.

A central task for mapping onto a classical Heisenberg model is the determination of exchange constants and magnetocrystalline
anisotropy. The exchange constants can be extracted from DFT calculations either by comparing the total energies of several
artificial collinear magnetic structures or by applying the magnetic force theorem in the framework of the Korringa-Kohn-Rostoker (KKR) Green's function
method\cite{gun,lic}. In addition to these two methods artificial noncollinear structures can be used to study exchange interactions in the Heisenberg model by choosing noncolinear states that can be controllably switched on and off. Noncollinear configurations used to calculate the exchange parameters for free-standing and embedded wires are given in Refs.[\onlinecite{hashemijap,hashemiphysica}].

\emph{Ab initio}  investigations of the freestanding Fe wire revealed that nearest neighbor exchange interactions dominate. \cite{hashemijap,hashemiphysica}
Next-nearest neighbor interactions are an order of magnitude smaller than smaller than nearest-neighbor interactions, therefore, we restrict the Heisenberg model to nearest neighbor interactions only.
Exchange constants $J_{ij}$ can be extracted from DFT calculations by comparing the total energies of several artificial noncollinear magnetic structures. In this approach we selectively switch on or off interactions between atoms $i$ and $j$ by deliberately choosing those noncollinear states.

\begin{figure}
\centering
\includegraphics[trim = 1.8cm 4cm 11cm 3.8cm, clip,scale=0.180, angle=0]{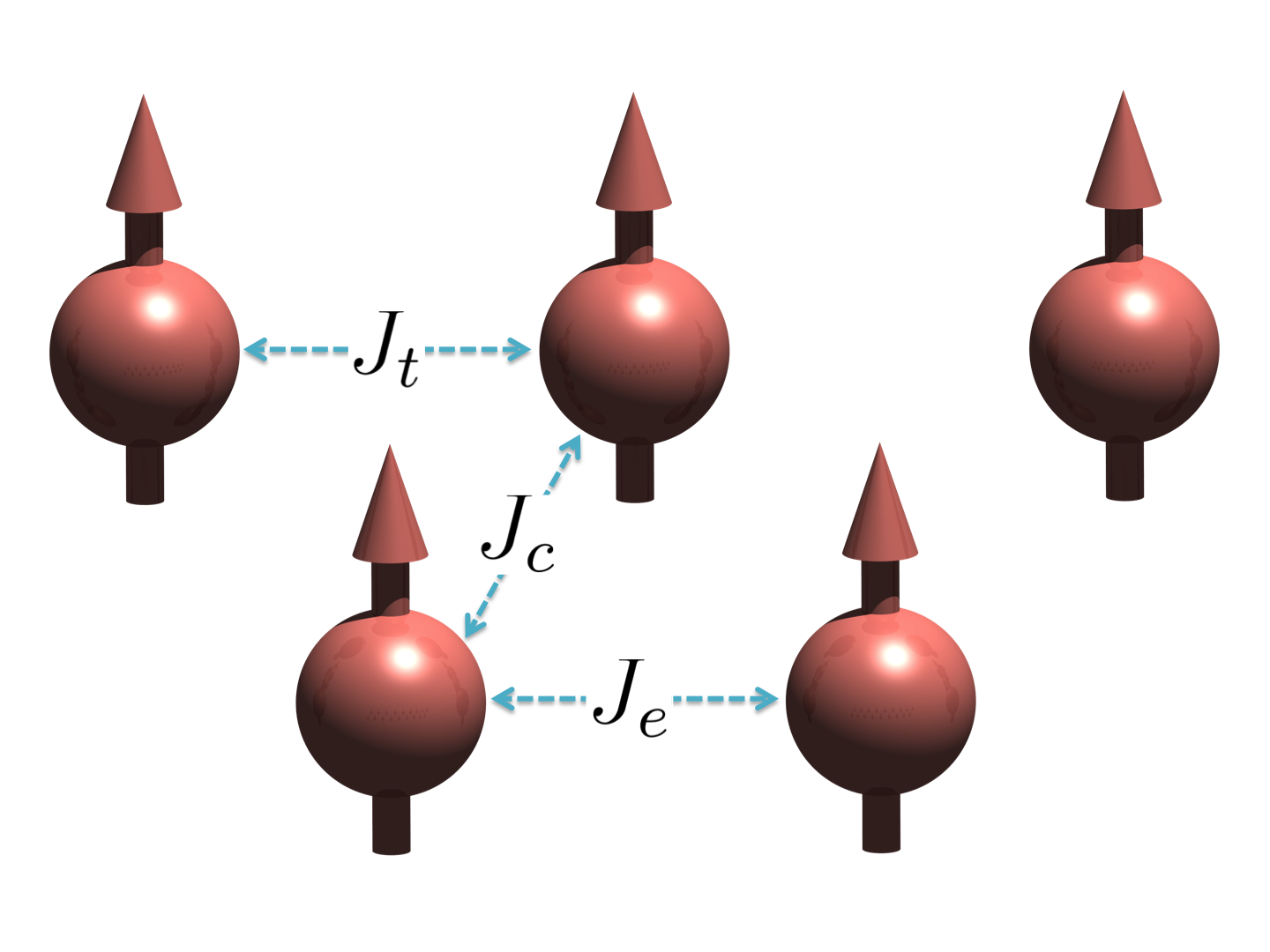}
\caption{ \label{fig:j} Schematic picture showing the Fe wires have the $\textrm{FF}$ ground state and also the magnetic interactions within the Fe wires; $J_{t}$, $J_{c}$, and $J_{e}$. }
\end{figure}

$J_{\parallel}$ can be broken down into three main couplings (as shown in Fig.\ref{fig:j}) : magnetic couplings between Fe atoms at the top position, $J_{t}$, crossing magnetic couplings between the Fe chain at the
 top position and the embedded Fe chain, $J_{c}$, and magnetic coupling between embedded Fe atoms, $J_{e}$.

Noncollinear configurations used for calculation of the exchange parameters for freestanding Fe chain, freestanding and embedded Fe wires can be found in Ref. [\onlinecite{hashemijap}]. The obtained exchange parameters published in Ref. [\onlinecite{hashemijap}] for Fe systems are summarized in Tab. \ref{tab:1}. As depicted in Fig.\ref{fig:j}, the top and the embedded Fe moments are ferromagnetically ordered.\cite{hashemiprb}

Our calculations show that the magnetic moments are constant for the different noncollinear configurations within a specific system and symmetrically equivalent arrangements  lead to the same exchange parameters. These data suggest that the exchange constant for free-standing Fe chain is consistent with the literature. \cite{tung, mokrousov,hashemiprb} The magnetic groundstate for free-standing Fe chain is in agreement with the results in Ref. [\onlinecite{tung}] for the relaxed chain since the relaxed bond length in the ferromagnetic state is close to the Cu bond length.

The calculated exchange constants show that Fe wires have ferromagnetic ground states and relaxation effects are seen in the exchange constants of the embedded systems. The exchange constants in Fe wire are generally smaller than in corresponding linear Fe chains due to an increased coordination number in the planar equilateral triangle ribbon form of Fe wires. The stronger hybridization due to the inward relaxation of the Fe wires leads to smaller intrawire exchange constants.


%

%
%
\begin{table}[h!]
\caption{\label{tab:1} Exchange constants for the freestanding Fe
chain, freestanding Fe wires, and the embedded Fe wires.
The definition of the constants in the Heisenberg model
incorporates the magnetic spin moments. All values are given in meV.}
\begin{ruledtabular}
 \begin{tabular}{ccc}
\vspace*{-6pt}\\
\multicolumn{3}{c}{\textit{freestanding Fe chain}} \rule[0pt]{0pt}{5pt} \\
\hline
$J$ &  114.54  \\
\hline
\vspace*{-6pt}\\
\multicolumn{3}{c}{\textit{freestanding Fe wire}} \rule[0pt]{0pt}{5pt} \\
\hline
$J_{t}$ &  80.33  \\
$J_{c}$ &150.42  \\
$J_{e}$ &   80.33  \\
\hline
\vspace*{-6pt}\\
\multicolumn{3}{c}{\textit{embedded Fe wire}}\rule[0pt]{0pt}{5pt}  \\
\hline
$J_{t}$ &79.76\\
$J_{c}$ & 80.00 \\
$J_{e}$ & 74.58
 \end{tabular}
\end{ruledtabular}
\end{table}

\subsubsection{$J_{\perp}$: Interwire exchange coupling \label{sec:Interwireexchangecoupling}}

\begin{figure}
\centering
\includegraphics[trim = 0cm 0cm 0cm 0cm, clip, scale=0.3, angle=0]{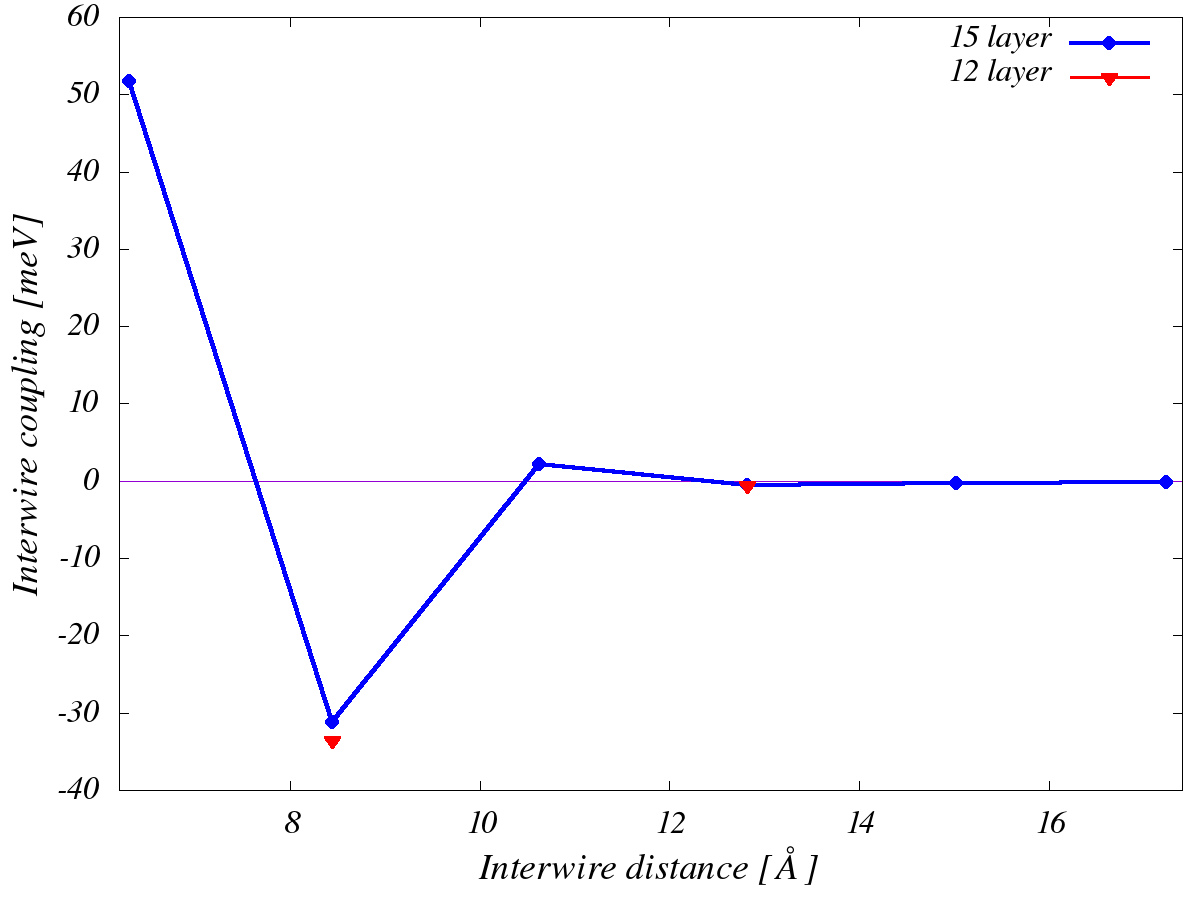}
\caption{The interwire coupling constants of Fe wires as functions of interwire separation. Blue square and red triangle symbols represents the couplings calculated using 12 and 15 layer slabs, respectively. The blue line connecting the symbols is a guide to the eye. \label{fig:exchange}}
\end{figure}

It is known that surface-state electrons on the (111) surfaces of noble metals create a two-dimensional (2D) nearly free electron gas which are confined to top layers at the surface. Electrons in these states move along the surface causing scattering of the surface electrons by nanostructures formed on the surface. This scattering leads to quantum interference patterns in the local density of states (LDOS) and long-range oscillatory interactions between adsorbates. \cite{Blongrange} In previous studies, the long range interactions have been attributed to the surface states of Cu(111) surfaces.\cite{PhysRevLett.98.056601,Stepanyuk2006272, PhysRevB.76.033409} Based on the calculations presented in Refs.[\onlinecite{hashemithesis}], we concluded that only considering 18 layer slabs of Cu produces a band structure which is comparable to experimental energy dispersion. Here, the interwire coupling constants are determined by making parallel and antiparallel alignments of the moments in the Fe wires: energy differences are calculated not according to the absolute value of energy alone. Our calculations showed that the interwire couplings (energy difference) converge faster than the absolute values of energies. These couplings were converged for slabs as thin as 12 layers. As one can see in Fig. \ref{fig:exchange}, using a 15 or 12 layer slab of Cu, gives a negligible difference for the interwire couplings, obtained for two interwire separations. Therefore, a 12 layer slab of Cu is used to simulate the Cu(111) surface and its surface states. A direct relaxation calculation for such a big system is very expensive, therefor the four top most relaxed layers of an eight layer slab of Cu(111) have been taken and replaced on the corresponding geometry of a twelve layer slab of Cu(111) surface, mimicking the relaxed geometry while the eight remaining bottom layers are fixed in their ideal bulk positions. The strength of the interwire magnetic coupling can be deduced from the energy difference of the ferro-magnetic and antiferromagnetic oriented wires, with supercells doubled in a direction perpendicular to the wires.

To construct a Heisenberg Hamiltonian which takes into account all the magnetic interactions, $J_{\parallel}$, effective intrawire couplings and $J_{\perp}$,  the interwire couplings are required. The effective intrawire couplings were determined in Sec. \ref{sec:Intrawireexchangecoupling}.  Now, we discuss how to estimate the interwire couplings. In principle, there are three interchain coupling constants in the cell. The first is the coupling between the embedded Fe chain and embedded Fe chain in the nearest neighboring wire. The second is the coupling between the embedded Fe chain and the deposited Fe chain at the top position in the nearest neighboring wire. Finally, there is the coupling between the deposited Fe chain on top and the deposited Fe chain at the top position in the nearest neighboring wire. These calculations are computationally demanding due to possible magnetic configurations besides ferromagnetic and antiferromagnetic. These extra configurations are necessary for as correction factors for exchange couplings even though they may not be energetically favored. The difference between the calculated exchange interactions may be too small, which begs the question whether this difference has a significant effect on the estimated transition temperature. Therefore the current study has been confined to effective interwire coupling constants. The interwire coupling constants were estimated using Equ. (\ref{sec:Intrawireexchangecoupling})  by making parallel and antiparallel alignment of the moments on the two wires on each side of the supercell.

The calculated exchange couplings as a function of interwire separation are shown in Fig. \ref{fig:exchange} and summarized in Tab. \ref{tab:jperp}. The exchange constants reflect the result that wires have a ferromagnetic groundstate on Cu(422). But there is a strong antiferromagnetic coupling for the nanowires on Cu(533). On Cu(644) the coupling becomes weakly ferromagnetic again. However on higher-index Cu(111) vicinal surfaces, we observe weak antiferromagnetic ordering. Relaxation effects are insignificant for the interwire couplings.

Similar to the current study, RKKY interactions have been observed not only in metallic layered systems but also between magnetic nanostructures deposited on metal surfaces in which the magnetic interactions are often mediated by surface-state electrons.\cite{Khajetoorians1062,PhysRevB.85.045429,PhysRevB.83.224416, PhysRevB.78.165413}

A rough estimation of the envelope of the magnitude of $J_{\perp}$ suggests an asymptotic decay with the inverse square of the interwire separation. This is in agreement with \emph{ab initio} calculations and STM experiments that predicted similar interactions between 3$d$ magnetic nanostructures on a Cu(111) surface caused by surface-state electrons. \cite{PhysRevB.68.205410, PhysRevB.70.075414,PhysRevB.83.224416} It is worth noting that the Cu bulk states can affect the interaction energies at relatively short interwire separation because the Fe wires couple to the Cu bulk bands as well. The magnetic interaction energy in the bulk asymptotically decays as the inverse fifth power of the interatomic distance.\cite{LAU197869}

Long-range interactions between the nanostructures on vicinal Cu(111) surfaces are distinct from those on Cu(111) surface for two reasons: First, the surfaces-states are affected by the electronic potential at the steps which is close to the Fermi energy in vicinal Cu(111) surfaces. Second, The Fe wires significantly affect the surface-states on vicinal surfaces.\cite{PhysRevB.75.155428}

\begin{table}[h!]
\caption{Exchange coupling constants ($J_{\perp}$) of Fe wires on Cu(111) stepped surfaces.   \label{tab:jperp}}
\begin{ruledtabular}
 \begin{tabular}{lcccccc}
Surface               &  (4,2,2)     & (5,3,3)    & (6,4,4)   & (7,5,5)    & (8,6,6) & (9,7,7)  \\
\hline
Seperation (\AA)      &  6.31        & 8.45       & 10.62     & 12.81      & 15.02   & 17.23    \\
\hline
$J_{\perp}$ (meV)     &  +51.75      & -33.52     & +2.16     & -0.55      &  -0.32   &  -0.14  \\
 \end{tabular}
\end{ruledtabular}
\end{table}

\subsection{Magnetism at Finite Temperatures\label{sec:MagnetismatFiniteTemperatures}}

It is necessary to calculate some well-defined macroscopic property which ensures the correct implementation of interactions in a system. Critical temperature ($T_c$) of the investigated systems is determined using Monte Carlo simulations. $T_c$ of a nanowire is primarily determined by the strength
of the exchange interaction between spins and the magnetic anisotropy energies. For the representation of the interwire interaction, $J_{\perp}$ is used as given in Tab. \ref{tab:jperp}.

During Monte Carlo simulations, lattices with 40$\times$40, 42$\times$42, and 44$\times$44 unit cells with 4 atoms per unit cell are used. The MAE of the first row of the Fe the Fe sublattices are equal.\cite{hashemianisotropy} Periodic boundary conditions are applied in the direction of and perpendicular to the wires. The critical temperature of the system is found  by relaxing into thermodynamical equilibrium with 10,000 MC steps per temperature step. Correlation effects are accounted for by averaging over 15,000 measurements, between each of which are two MC steps. To improve the statistics, averaging over 20 temperature loops is performed and importance sampling is performed using the Metropolis algorithm. $T_c$$'$s are determined using the specific heat and, in the case of a ferromagnetic system, the susceptibility $\chi$ and 4th-order-cumulant $U_4$.

\begin{figure}
\begin{center}
\centering
\includegraphics[scale=0.26]{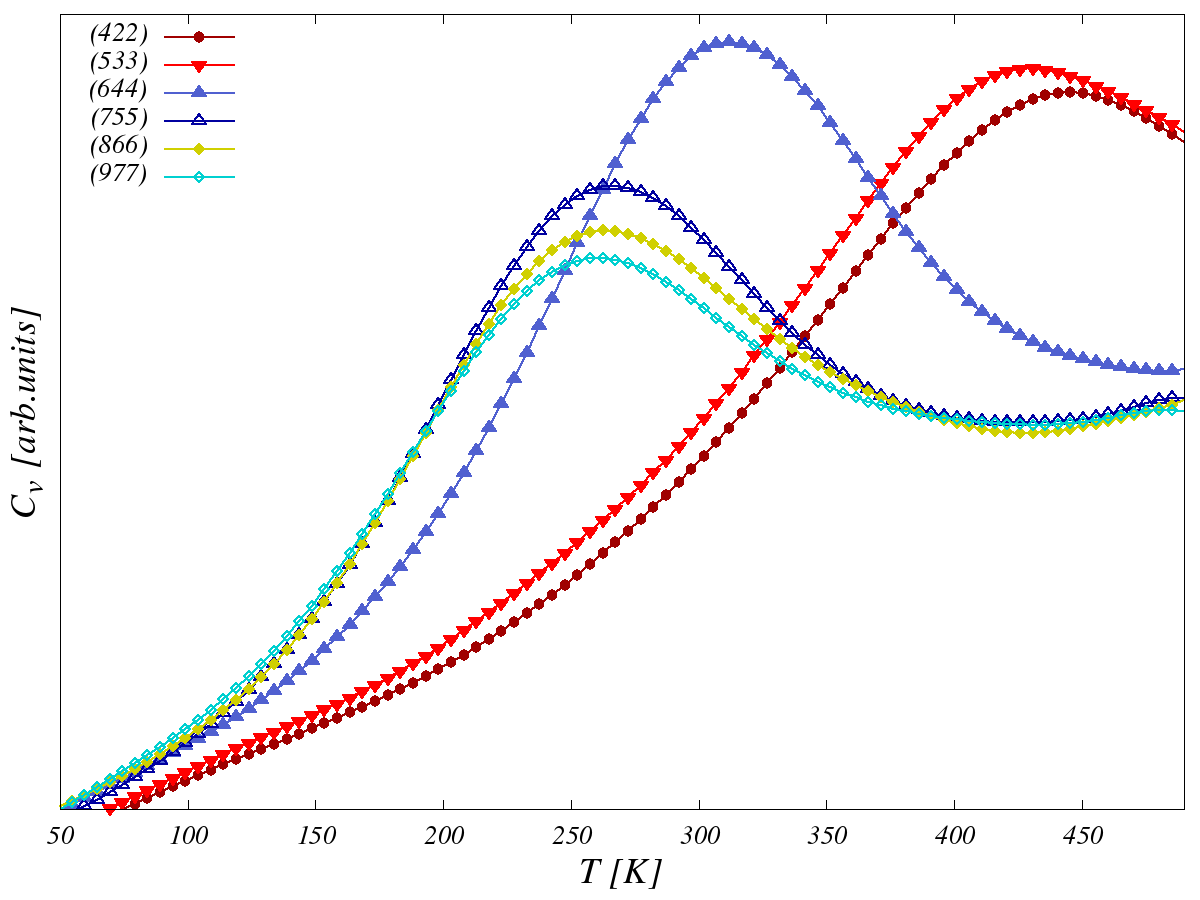}
\caption{\label{fig:Specificheat} Heat capacity of the Fe nanowires for the MAE of 4.76 meV.}
\end{center}
\end{figure}

Figure\ref{fig:Specificheat} shows the specific heats for a Fe wire on the vicinal Cu(111) surfaces. We observe that there is a $T_c$ for all systems. In our previous study,\cite{hashemijap} we observed no magnetic ordering for a vanishing MAE, in agreement with the Mermin--Wagner theorem.\cite{PhysRevLett.17.1133} Our results also show that an increase in interwire couplings stabilizes the moments against thermal fluctuation and, thus, leads to an increase of the $T_c$. Furthermore, the phase transition broadens as the interwire couplings are increased. The calculated $T_c$$'$s are below room temperature and close to each other for Fe wires on Cu(977), Cu(866), and Cu(755). The interwire couplings $J_{\perp}$ are very small and the $T_c$$'$s for these systems are determined their intrawire couplings,  which  are  the  interactions responsible for the ferromagnetic ordering within the Fe wires. The $T_c$ is close to room temperature for Fe wire on Cu(644) and well-above room temperature for Cu(533) and C(422). This can be traced  back  to  the  high  values of $J_{\perp}$ for these systems. A summary of the $T_c$$'$s of all systems for MAE=4.76 meV can be found in Tab. \ref{tab:Tc}.

\begin{table}[h!]
	\caption{Critical temperatures ($T_c$) of Fe wires on the vicinal Cu(111) surfaces for the MAE of 4.76 meV. The temperature values are given in Kelvin.\label{tab:Tc}}
	\begin{ruledtabular}
		\begin{tabular}{ccccccc}
			Surface       &  (4,2,2)     & (5,3,3)    & (6,4,4)   & (7,5,5)    & (8,6,6) & (9,7,7)  \\
			\hline
			$T_{c}$       &  445        & 431         & 312       & 271        & 266    &   262     \\
		\end{tabular}
	\end{ruledtabular}
\end{table}

\section{Conclusions\label{sec:Conclusions}}

\textit{Ab initio} DFT calculations have been used to set up a classical anisotropic Heisenberg model to study finite temperature properties of Fe wires embedded in Cu(111). To do this, exchange parameters including intrawire and interwire couplings are extracted from  non-collinear and collinear DFT calculations, respectively. The intrawire couplings are one order of magnitude higher than the interwire couplings, for relatively large interwire separation. A slab of of Cu at least 12 layers thick is used to simulate the Cu(111) surface states to provide converged values for the interwire couplings. The interwire exchange couplings of the Fe wires across the vicinal Cu(111) surface oscillate with the interwire separation. This provides reliable means for stabilizing the magnetic ordering of the Fe nanowires in either a ferromagnetic or an antiferromagnetic configuration.

This study provides a technologically feasible way of tailoring 1D magnetic nanostructures adsorbed on a vicinal Cu(111) surface. The critical temperatures of the systems with shorter interwire separation is well-above room temperature. This is a strong indication that these nanowires have potential applications in high-density magnetic data storages.

\noindent \textbf{Acknowledgement}\\
The work was supported by the cluster of excellence
Nanostructured Materials of the state Saxony-Anhalt and the
International Max Planck Research School for Science and
Technology of Nanostructures. We acknowledge the resources
provided by the University of Michigan’s Advanced Research Computing.

\bibliography{exchange}

\begin{thebibliography}{43}%
\makeatletter
\providecommand \@ifxundefined [1]{%
 \@ifx{#1\undefined}
}%
\providecommand \@ifnum [1]{%
 \ifnum #1\expandafter \@firstoftwo
 \else \expandafter \@secondoftwo
 \fi
}%
\providecommand \@ifx [1]{%
 \ifx #1\expandafter \@firstoftwo
 \else \expandafter \@secondoftwo
 \fi
}%
\providecommand \natexlab [1]{#1}%
\providecommand \enquote  [1]{``#1''}%
\providecommand \bibnamefont  [1]{#1}%
\providecommand \bibfnamefont [1]{#1}%
\providecommand \citenamefont [1]{#1}%
\providecommand \href@noop [0]{\@secondoftwo}%
\providecommand \href [0]{\begingroup \@sanitize@url \@href}%
\providecommand \@href[1]{\@@startlink{#1}\@@href}%
\providecommand \@@href[1]{\endgroup#1\@@endlink}%
\providecommand \@sanitize@url [0]{\catcode `\\12\catcode `\$12\catcode
  `\&12\catcode `\#12\catcode `\^12\catcode `\_12\catcode `\%12\relax}%
\providecommand \@@startlink[1]{}%
\providecommand \@@endlink[0]{}%
\providecommand \url  [0]{\begingroup\@sanitize@url \@url }%
\providecommand \@url [1]{\endgroup\@href {#1}{\urlprefix }}%
\providecommand \urlprefix  [0]{URL }%
\providecommand \Eprint [0]{\href }%
\providecommand \doibase [0]{http://dx.doi.org/}%
\providecommand \selectlanguage [0]{\@gobble}%
\providecommand \bibinfo  [0]{\@secondoftwo}%
\providecommand \bibfield  [0]{\@secondoftwo}%
\providecommand \translation [1]{[#1]}%
\providecommand \BibitemOpen [0]{}%
\providecommand \bibitemStop [0]{}%
\providecommand \bibitemNoStop [0]{.\EOS\space}%
\providecommand \EOS [0]{\spacefactor3000\relax}%
\providecommand \BibitemShut  [1]{\csname bibitem#1\endcsname}%
\let\auto@bib@innerbib\@empty
\bibitem [{\citenamefont {Shen}\ \emph
  {et~al.}(1997{\natexlab{a}})\citenamefont {Shen}, \citenamefont {Skomski},
  \citenamefont {Klaua}, \citenamefont {Jenniches}, \citenamefont {Manoharan},\
  and\ \citenamefont {Kirschner}}]{shen1}%
  \BibitemOpen
  \bibfield  {author} {\bibinfo {author} {\bibfnamefont {J.}~\bibnamefont
  {Shen}}, \bibinfo {author} {\bibfnamefont {R.}~\bibnamefont {Skomski}},
  \bibinfo {author} {\bibfnamefont {M.}~\bibnamefont {Klaua}}, \bibinfo
  {author} {\bibfnamefont {H.}~\bibnamefont {Jenniches}}, \bibinfo {author}
  {\bibfnamefont {S.~S.}\ \bibnamefont {Manoharan}}, \ and\ \bibinfo {author}
  {\bibfnamefont {J.}~\bibnamefont {Kirschner}},\ }\href {\doibase
  10.1103/PhysRevB.56.2340} {\bibfield  {journal} {\bibinfo  {journal} {Phys.
  Rev. B}\ }\textbf {\bibinfo {volume} {56}},\ \bibinfo {pages} {2340}
  (\bibinfo {year} {1997}{\natexlab{a}})}\BibitemShut {NoStop}%
\bibitem [{\citenamefont {Shen}\ \emph
  {et~al.}(1997{\natexlab{b}})\citenamefont {Shen}, \citenamefont {Klaua},
  \citenamefont {Ohresser}, \citenamefont {Jenniches}, \citenamefont {Barthel},
  \citenamefont {Mohan},\ and\ \citenamefont {Kirschner}}]{shen2}%
  \BibitemOpen
  \bibfield  {author} {\bibinfo {author} {\bibfnamefont {J.}~\bibnamefont
  {Shen}}, \bibinfo {author} {\bibfnamefont {M.}~\bibnamefont {Klaua}},
  \bibinfo {author} {\bibfnamefont {P.}~\bibnamefont {Ohresser}}, \bibinfo
  {author} {\bibfnamefont {H.}~\bibnamefont {Jenniches}}, \bibinfo {author}
  {\bibfnamefont {J.}~\bibnamefont {Barthel}}, \bibinfo {author} {\bibfnamefont
  {C.~V.}\ \bibnamefont {Mohan}}, \ and\ \bibinfo {author} {\bibfnamefont
  {J.}~\bibnamefont {Kirschner}},\ }\href {\doibase 10.1103/PhysRevB.56.11134}
  {\bibfield  {journal} {\bibinfo  {journal} {Phys. Rev. B}\ }\textbf {\bibinfo
  {volume} {56}},\ \bibinfo {pages} {11134} (\bibinfo {year}
  {1997}{\natexlab{b}})}\BibitemShut {NoStop}%
\bibitem [{\citenamefont {Guo}\ \emph {et~al.}(2006)\citenamefont {Guo},
  \citenamefont {Mo}, \citenamefont {Kaxiras}, \citenamefont {Zhang},\ and\
  \citenamefont {Weitering}}]{mo2}%
  \BibitemOpen
  \bibfield  {author} {\bibinfo {author} {\bibfnamefont {J.}~\bibnamefont
  {Guo}}, \bibinfo {author} {\bibfnamefont {Y.}~\bibnamefont {Mo}}, \bibinfo
  {author} {\bibfnamefont {E.}~\bibnamefont {Kaxiras}}, \bibinfo {author}
  {\bibfnamefont {Z.}~\bibnamefont {Zhang}}, \ and\ \bibinfo {author}
  {\bibfnamefont {H.~H.}\ \bibnamefont {Weitering}},\ }\href@noop {} {\bibfield
   {journal} {\bibinfo  {journal} {Phys. Rev. B}\ }\textbf {\bibinfo {volume}
  {73}},\ \bibinfo {eid} {193405} (\bibinfo {year} {2006})}\BibitemShut
  {NoStop}%
\bibitem [{\citenamefont {Hirjibehedin}\ \emph {et~al.}(2006)\citenamefont
  {Hirjibehedin}, \citenamefont {Lutz},\ and\ \citenamefont
  {Heinrich}}]{Cyrus}%
  \BibitemOpen
  \bibfield  {author} {\bibinfo {author} {\bibfnamefont {C.~F.}\ \bibnamefont
  {Hirjibehedin}}, \bibinfo {author} {\bibfnamefont {C.~P.}\ \bibnamefont
  {Lutz}}, \ and\ \bibinfo {author} {\bibfnamefont {A.~J.}\ \bibnamefont
  {Heinrich}},\ }\href {\doibase 10.1126/science.1125398} {\bibfield  {journal}
  {\bibinfo  {journal} {Science}\ }\textbf {\bibinfo {volume} {312}},\ \bibinfo
  {pages} {1021} (\bibinfo {year} {2006})}\BibitemShut {NoStop}%
\bibitem [{\citenamefont {Gambardella}\ \emph {et~al.}(2002)\citenamefont
  {Gambardella}, \citenamefont {Dallmeyer}, \citenamefont {Maiti},
  \citenamefont {Malagoli}, \citenamefont {Eberhardt}, \citenamefont {Kern},\
  and\ \citenamefont {Carbone}}]{Gambardella}%
  \BibitemOpen
  \bibfield  {author} {\bibinfo {author} {\bibfnamefont {P.}~\bibnamefont
  {Gambardella}}, \bibinfo {author} {\bibfnamefont {A.}~\bibnamefont
  {Dallmeyer}}, \bibinfo {author} {\bibfnamefont {K.}~\bibnamefont {Maiti}},
  \bibinfo {author} {\bibfnamefont {M.~C.}\ \bibnamefont {Malagoli}}, \bibinfo
  {author} {\bibfnamefont {W.}~\bibnamefont {Eberhardt}}, \bibinfo {author}
  {\bibfnamefont {K.}~\bibnamefont {Kern}}, \ and\ \bibinfo {author}
  {\bibfnamefont {C.}~\bibnamefont {Carbone}},\ }\href {\doibase
  10.1103/PhysRevB.77.214413} {\bibfield  {journal} {\bibinfo  {journal}
  {Nature}\ }\textbf {\bibinfo {volume} {416}},\ \bibinfo {eid} {301} (\bibinfo
  {year} {2002})}\BibitemShut {NoStop}%
\bibitem [{\citenamefont {Lounis}\ \emph {et~al.}(2008)\citenamefont {Lounis},
  \citenamefont {Dederichs},\ and\ \citenamefont {Bl\"{u}gel}}]{lds08}%
  \BibitemOpen
  \bibfield  {author} {\bibinfo {author} {\bibfnamefont {S.}~\bibnamefont
  {Lounis}}, \bibinfo {author} {\bibfnamefont {P.~H.}\ \bibnamefont
  {Dederichs}}, \ and\ \bibinfo {author} {\bibfnamefont {S.}~\bibnamefont
  {Bl\"{u}gel}},\ }\href@noop {} {\bibfield  {journal} {\bibinfo  {journal}
  {Phys. Rev. Lett.}\ }\textbf {\bibinfo {volume} {101}},\ \bibinfo {pages}
  {107204} (\bibinfo {year} {2008})}\BibitemShut {NoStop}%
\bibitem [{\citenamefont {Lazarovits}\ \emph {et~al.}(2003)\citenamefont
  {Lazarovits}, \citenamefont {Szunyogh}, \citenamefont {Weinberger},\ and\
  \citenamefont {\'Ujfalussy}}]{lsw03}%
  \BibitemOpen
  \bibfield  {author} {\bibinfo {author} {\bibfnamefont {B.}~\bibnamefont
  {Lazarovits}}, \bibinfo {author} {\bibfnamefont {L.}~\bibnamefont
  {Szunyogh}}, \bibinfo {author} {\bibfnamefont {P.}~\bibnamefont
  {Weinberger}}, \ and\ \bibinfo {author} {\bibfnamefont {B.}~\bibnamefont
  {\'Ujfalussy}},\ }\href@noop {} {\bibfield  {journal} {\bibinfo  {journal}
  {Phys. Rev. B}\ }\textbf {\bibinfo {volume} {68}},\ \bibinfo {pages} {024433}
  (\bibinfo {year} {2003})}\BibitemShut {NoStop}%
\bibitem [{\citenamefont {Tung}\ and\ \citenamefont {Guo}(2007)}]{tung}%
  \BibitemOpen
  \bibfield  {author} {\bibinfo {author} {\bibfnamefont {J.~C.}\ \bibnamefont
  {Tung}}\ and\ \bibinfo {author} {\bibfnamefont {G.~Y.}\ \bibnamefont {Guo}},\
  }\href {\doibase 10.1103/PhysRevB.76.094413} {\bibfield  {journal} {\bibinfo
  {journal} {Phys. Rev. B}\ }\textbf {\bibinfo {volume} {76}},\ \bibinfo {eid}
  {094413} (\bibinfo {year} {2007})}\BibitemShut {NoStop}%
\bibitem [{\citenamefont {Mokrousov}\ \emph
  {et~al.}(2007{\natexlab{a}})\citenamefont {Mokrousov}, \citenamefont
  {Bihlmayer}, \citenamefont {Bl\"{u}gel},\ and\ \citenamefont
  {Heinze}}]{mbb07}%
  \BibitemOpen
  \bibfield  {author} {\bibinfo {author} {\bibfnamefont {Y.}~\bibnamefont
  {Mokrousov}}, \bibinfo {author} {\bibfnamefont {G.}~\bibnamefont
  {Bihlmayer}}, \bibinfo {author} {\bibfnamefont {S.}~\bibnamefont
  {Bl\"{u}gel}}, \ and\ \bibinfo {author} {\bibfnamefont {S.}~\bibnamefont
  {Heinze}},\ }\href@noop {} {\bibfield  {journal} {\bibinfo  {journal} {Phys.
  Rev. B}\ }\textbf {\bibinfo {volume} {75}},\ \bibinfo {pages} {104413}
  (\bibinfo {year} {2007}{\natexlab{a}})}\BibitemShut {NoStop}%
\bibitem [{\citenamefont {Mo}\ \emph {et~al.}(2008)\citenamefont {Mo},
  \citenamefont {Zhu}, \citenamefont {Kaxiras},\ and\ \citenamefont
  {Zhang}}]{mzk08}%
  \BibitemOpen
  \bibfield  {author} {\bibinfo {author} {\bibfnamefont {Y.}~\bibnamefont
  {Mo}}, \bibinfo {author} {\bibfnamefont {W.}~\bibnamefont {Zhu}}, \bibinfo
  {author} {\bibfnamefont {E.}~\bibnamefont {Kaxiras}}, \ and\ \bibinfo
  {author} {\bibfnamefont {Z.}~\bibnamefont {Zhang}},\ }\href@noop {}
  {\bibfield  {journal} {\bibinfo  {journal} {Phys. Rev. Lett.}\ }\textbf
  {\bibinfo {volume} {101}},\ \bibinfo {pages} {216101} (\bibinfo {year}
  {2008})}\BibitemShut {NoStop}%
\bibitem [{\citenamefont {Spi\ifmmode~\check{s}\else \v{s}\fi{}\'ak}\ and\
  \citenamefont {Hafner}(2002)}]{sh02}%
  \BibitemOpen
  \bibfield  {author} {\bibinfo {author} {\bibfnamefont {D.}~\bibnamefont
  {Spi\ifmmode~\check{s}\else \v{s}\fi{}\'ak}}\ and\ \bibinfo {author}
  {\bibfnamefont {J.}~\bibnamefont {Hafner}},\ }\href@noop {} {\bibfield
  {journal} {\bibinfo  {journal} {Phy. Rev. B}\ }\textbf {\bibinfo {volume}
  {65}},\ \bibinfo {pages} {235405} (\bibinfo {year} {2002})}\BibitemShut
  {NoStop}%
\bibitem [{\citenamefont {Spi\ifmmode~\check{s}\else \v{s}\fi{}\'ak}\ and\
  \citenamefont {Hafner}(2003{\natexlab{a}})}]{sh03}%
  \BibitemOpen
  \bibfield  {author} {\bibinfo {author} {\bibfnamefont {D.}~\bibnamefont
  {Spi\ifmmode~\check{s}\else \v{s}\fi{}\'ak}}\ and\ \bibinfo {author}
  {\bibfnamefont {J.}~\bibnamefont {Hafner}},\ }\href@noop {} {\bibfield
  {journal} {\bibinfo  {journal} {Phys. Rev. B}\ }\textbf {\bibinfo {volume}
  {67}},\ \bibinfo {pages} {214416} (\bibinfo {year}
  {2003}{\natexlab{a}})}\BibitemShut {NoStop}%
\bibitem [{\citenamefont {Spi\ifmmode~\check{s}\else \v{s}\fi{}\'ak}\ and\
  \citenamefont {Hafner}(2003{\natexlab{b}})}]{sh03a}%
  \BibitemOpen
  \bibfield  {author} {\bibinfo {author} {\bibfnamefont {D.}~\bibnamefont
  {Spi\ifmmode~\check{s}\else \v{s}\fi{}\'ak}}\ and\ \bibinfo {author}
  {\bibfnamefont {J.}~\bibnamefont {Hafner}},\ }\href@noop {} {\bibfield
  {journal} {\bibinfo  {journal} {Comput. Mater. Sci.}\ }\textbf {\bibinfo
  {volume} {27}},\ \bibinfo {pages} {138} (\bibinfo {year}
  {2003}{\natexlab{b}})}\BibitemShut {NoStop}%
\bibitem [{\citenamefont {Spi\ifmmode~\check{s}\else \v{s}\fi{}\'ak}\ and\
  \citenamefont {Hafner}(2004)}]{sh04a}%
  \BibitemOpen
  \bibfield  {author} {\bibinfo {author} {\bibfnamefont {D.}~\bibnamefont
  {Spi\ifmmode~\check{s}\else \v{s}\fi{}\'ak}}\ and\ \bibinfo {author}
  {\bibfnamefont {J.}~\bibnamefont {Hafner}},\ }\href@noop {} {\bibfield
  {journal} {\bibinfo  {journal} {Comput. Mater. Sci.}\ }\textbf {\bibinfo
  {volume} {30}},\ \bibinfo {pages} {278} (\bibinfo {year} {2004})}\BibitemShut
  {NoStop}%
\bibitem [{\citenamefont {Mo}\ \emph {et~al.}(2005)\citenamefont {Mo},
  \citenamefont {Varga}, \citenamefont {Kaxiras},\ and\ \citenamefont
  {Zhang}}]{mo1}%
  \BibitemOpen
  \bibfield  {author} {\bibinfo {author} {\bibfnamefont {Y.}~\bibnamefont
  {Mo}}, \bibinfo {author} {\bibfnamefont {K.}~\bibnamefont {Varga}}, \bibinfo
  {author} {\bibfnamefont {E.}~\bibnamefont {Kaxiras}}, \ and\ \bibinfo
  {author} {\bibfnamefont {Z.}~\bibnamefont {Zhang}},\ }\href@noop {}
  {\bibfield  {journal} {\bibinfo  {journal} {Phys. Rev. Lett.}\ }\textbf
  {\bibinfo {volume} {94}},\ \bibinfo {eid} {155503} (\bibinfo {year}
  {2005})}\BibitemShut {NoStop}%
\bibitem [{\citenamefont {Tegenkamp}(2009)}]{teg09}%
  \BibitemOpen
  \bibfield  {author} {\bibinfo {author} {\bibfnamefont {C.}~\bibnamefont
  {Tegenkamp}},\ }\href@noop {} {\bibfield  {journal} {\bibinfo  {journal} {J.
  Phys.: Condens. Matter}\ }\textbf {\bibinfo {volume} {21}},\ \bibinfo {pages}
  {1013002} (\bibinfo {year} {2009})}\BibitemShut {NoStop}%
\bibitem [{\citenamefont {Negulyaev}\ \emph {et~al.}(2008)\citenamefont
  {Negulyaev}, \citenamefont {Stepanyuk}, \citenamefont {Hergert},
  \citenamefont {Bruno},\ and\ \citenamefont {Kirschner}}]{negulyaev2008}%
  \BibitemOpen
  \bibfield  {author} {\bibinfo {author} {\bibfnamefont {N.~N.}\ \bibnamefont
  {Negulyaev}}, \bibinfo {author} {\bibfnamefont {V.~S.}\ \bibnamefont
  {Stepanyuk}}, \bibinfo {author} {\bibfnamefont {W.}~\bibnamefont {Hergert}},
  \bibinfo {author} {\bibfnamefont {P.}~\bibnamefont {Bruno}}, \ and\ \bibinfo
  {author} {\bibfnamefont {J.}~\bibnamefont {Kirschner}},\ }\href@noop {}
  {\bibfield  {journal} {\bibinfo  {journal} {Phy. Rev. B}\ }\textbf {\bibinfo
  {volume} {77}},\ \bibinfo {eid} {085430} (\bibinfo {year}
  {2008})}\BibitemShut {NoStop}%
\bibitem [{\citenamefont {Hashemi}\ \emph
  {et~al.}(2010{\natexlab{a}})\citenamefont {Hashemi}, \citenamefont
  {Hergert},\ and\ \citenamefont {Stepanyuk}}]{hashemiprb}%
  \BibitemOpen
  \bibfield  {author} {\bibinfo {author} {\bibfnamefont {H.}~\bibnamefont
  {Hashemi}}, \bibinfo {author} {\bibfnamefont {W.}~\bibnamefont {Hergert}}, \
  and\ \bibinfo {author} {\bibfnamefont {V.~S.}\ \bibnamefont {Stepanyuk}},\
  }\href {\doibase {10.1103/PhysRevB.81.104418}} {\bibfield  {journal}
  {\bibinfo  {journal} {Phys. Rev. B}\ }\textbf {\bibinfo {volume} {{81}}}
  (\bibinfo {year} {{2010}}{\natexlab{a}}),\
  {10.1103/PhysRevB.81.104418}}\BibitemShut {NoStop}%
\bibitem [{\citenamefont {Hashemi}\ \emph
  {et~al.}(2010{\natexlab{b}})\citenamefont {Hashemi}, \citenamefont
  {Hergert},\ and\ \citenamefont {Stepanyuk}}]{hashemijmmm}%
  \BibitemOpen
  \bibfield  {author} {\bibinfo {author} {\bibfnamefont {H.}~\bibnamefont
  {Hashemi}}, \bibinfo {author} {\bibfnamefont {W.}~\bibnamefont {Hergert}}, \
  and\ \bibinfo {author} {\bibfnamefont {V.}~\bibnamefont {Stepanyuk}},\ }\href
  {\doibase https://doi.org/10.1016/j.jmmm.2009.04.048} {\bibfield  {journal}
  {\bibinfo  {journal} {Journal of Magnetism and Magnetic Materials}\ }\textbf
  {\bibinfo {volume} {322}},\ \bibinfo {pages} {1296 } (\bibinfo {year}
  {2010}{\natexlab{b}})},\ \bibinfo {note} {proceedings of the Joint European
  Magnetic Symposia}\BibitemShut {NoStop}%
\bibitem [{\citenamefont {Kresse}\ and\ \citenamefont {Hafner}(1993)}]{vasp1}%
  \BibitemOpen
  \bibfield  {author} {\bibinfo {author} {\bibfnamefont {G.}~\bibnamefont
  {Kresse}}\ and\ \bibinfo {author} {\bibfnamefont {J.}~\bibnamefont
  {Hafner}},\ }\href {\doibase 10.1103/PhysRevB.48.13115} {\bibfield  {journal}
  {\bibinfo  {journal} {Phys. Rev. B}\ }\textbf {\bibinfo {volume} {48}},\
  \bibinfo {pages} {13115} (\bibinfo {year} {1993})}\BibitemShut {NoStop}%
\bibitem [{\citenamefont {Kresse}\ and\ \citenamefont
  {Furthm\"uller}(1996)}]{vasp2}%
  \BibitemOpen
  \bibfield  {author} {\bibinfo {author} {\bibfnamefont {G.}~\bibnamefont
  {Kresse}}\ and\ \bibinfo {author} {\bibfnamefont {J.}~\bibnamefont
  {Furthm\"uller}},\ }\href {\doibase 10.1103/PhysRevB.54.11169} {\bibfield
  {journal} {\bibinfo  {journal} {Phys. Rev. B}\ }\textbf {\bibinfo {volume}
  {54}},\ \bibinfo {pages} {11169} (\bibinfo {year} {1996})}\BibitemShut
  {NoStop}%
\bibitem [{\citenamefont {Bl\"ochl}(1994)}]{paw1}%
  \BibitemOpen
  \bibfield  {author} {\bibinfo {author} {\bibfnamefont {P.~E.}\ \bibnamefont
  {Bl\"ochl}},\ }\href {\doibase 10.1103/PhysRevB.50.17953} {\bibfield
  {journal} {\bibinfo  {journal} {Phys. Rev. B}\ }\textbf {\bibinfo {volume}
  {50}},\ \bibinfo {pages} {17953} (\bibinfo {year} {1994})}\BibitemShut
  {NoStop}%
\bibitem [{\citenamefont {Perdew}\ \emph {et~al.}(1996)\citenamefont {Perdew},
  \citenamefont {Burke},\ and\ \citenamefont
  {Ernzerhof}}]{PhysRevLett.77.3865}%
  \BibitemOpen
  \bibfield  {author} {\bibinfo {author} {\bibfnamefont {J.~P.}\ \bibnamefont
  {Perdew}}, \bibinfo {author} {\bibfnamefont {K.}~\bibnamefont {Burke}}, \
  and\ \bibinfo {author} {\bibfnamefont {M.}~\bibnamefont {Ernzerhof}},\ }\href
  {\doibase 10.1103/PhysRevLett.77.3865} {\bibfield  {journal} {\bibinfo
  {journal} {Phys. Rev. Lett.}\ }\textbf {\bibinfo {volume} {77}},\ \bibinfo
  {pages} {3865} (\bibinfo {year} {1996})}\BibitemShut {NoStop}%
\bibitem [{\citenamefont {Hashemi}\ \emph {et~al.}(2016)\citenamefont
  {Hashemi}, \citenamefont {Bregman}, \citenamefont {Nabi},\ and\ \citenamefont
  {Kieffer}}]{hashemianisotropy}%
  \BibitemOpen
  \bibfield  {author} {\bibinfo {author} {\bibfnamefont {H.}~\bibnamefont
  {Hashemi}}, \bibinfo {author} {\bibfnamefont {A.}~\bibnamefont {Bregman}},
  \bibinfo {author} {\bibfnamefont {H.~S.}\ \bibnamefont {Nabi}}, \ and\
  \bibinfo {author} {\bibfnamefont {J.}~\bibnamefont {Kieffer}},\ }\href
  {\doibase 10.1039/C6RA23541G} {\bibfield  {journal} {\bibinfo  {journal} {RSC
  Adv.}\ }\textbf {\bibinfo {volume} {6}},\ \bibinfo {pages} {108948} (\bibinfo
  {year} {2016})}\BibitemShut {NoStop}%
\bibitem [{\citenamefont {Fischer}\ \emph {et~al.}(2009)\citenamefont
  {Fischer}, \citenamefont {D\"ane}, \citenamefont {Ernst}, \citenamefont
  {Bruno}, \citenamefont {L\"uders}, \citenamefont {Szotek}, \citenamefont
  {Temmerman},\ and\ \citenamefont {Hergert}}]{gun}%
  \BibitemOpen
  \bibfield  {author} {\bibinfo {author} {\bibfnamefont {G.}~\bibnamefont
  {Fischer}}, \bibinfo {author} {\bibfnamefont {M.}~\bibnamefont {D\"ane}},
  \bibinfo {author} {\bibfnamefont {A.}~\bibnamefont {Ernst}}, \bibinfo
  {author} {\bibfnamefont {P.}~\bibnamefont {Bruno}}, \bibinfo {author}
  {\bibfnamefont {M.}~\bibnamefont {L\"uders}}, \bibinfo {author}
  {\bibfnamefont {Z.}~\bibnamefont {Szotek}}, \bibinfo {author} {\bibfnamefont
  {W.}~\bibnamefont {Temmerman}}, \ and\ \bibinfo {author} {\bibfnamefont
  {W.}~\bibnamefont {Hergert}},\ }\href {\doibase 10.1103/PhysRevB.80.014408}
  {\bibfield  {journal} {\bibinfo  {journal} {Phys. Rev. B}\ }\textbf {\bibinfo
  {volume} {80}},\ \bibinfo {pages} {014408} (\bibinfo {year}
  {2009})}\BibitemShut {NoStop}%
\bibitem [{\citenamefont {Liechtenstein}\ \emph {et~al.}(1987)\citenamefont
  {Liechtenstein}, \citenamefont {Katsnelson}, \citenamefont {Antropov},\ and\
  \citenamefont {Gubanov}}]{lic}%
  \BibitemOpen
  \bibfield  {author} {\bibinfo {author} {\bibfnamefont {A.}~\bibnamefont
  {Liechtenstein}}, \bibinfo {author} {\bibfnamefont {M.}~\bibnamefont
  {Katsnelson}}, \bibinfo {author} {\bibfnamefont {V.}~\bibnamefont
  {Antropov}}, \ and\ \bibinfo {author} {\bibfnamefont {V.}~\bibnamefont
  {Gubanov}},\ }\href {\doibase http://dx.doi.org/10.1016/0304-8853(87)90721-9}
  {\bibfield  {journal} {\bibinfo  {journal} {Journal of Magnetism and Magnetic
  Materials}\ }\textbf {\bibinfo {volume} {67}},\ \bibinfo {pages} {65 }
  (\bibinfo {year} {1987})}\BibitemShut {NoStop}%
\bibitem [{\citenamefont {Hashemi}\ \emph
  {et~al.}(2010{\natexlab{c}})\citenamefont {Hashemi}, \citenamefont {Fischer},
  \citenamefont {Hergert},\ and\ \citenamefont {Stepanyuk}}]{hashemijap}%
  \BibitemOpen
  \bibfield  {author} {\bibinfo {author} {\bibfnamefont {H.}~\bibnamefont
  {Hashemi}}, \bibinfo {author} {\bibfnamefont {G.}~\bibnamefont {Fischer}},
  \bibinfo {author} {\bibfnamefont {W.}~\bibnamefont {Hergert}}, \ and\
  \bibinfo {author} {\bibfnamefont {V.~S.}\ \bibnamefont {Stepanyuk}},\ }\href
  {\doibase 10.1063/1.3368794} {\bibfield  {journal} {\bibinfo  {journal}
  {Journal of Applied Physics}\ }\textbf {\bibinfo {volume} {107}},\ \bibinfo
  {pages} {09E311} (\bibinfo {year} {2010}{\natexlab{c}})},\ \Eprint
  {http://arxiv.org/abs/https://doi.org/10.1063/1.3368794}
  {https://doi.org/10.1063/1.3368794} \BibitemShut {NoStop}%
\bibitem [{\citenamefont {Ignatiev}\ \emph {et~al.}(2010)\citenamefont
  {Ignatiev}, \citenamefont {Negulyaev}, \citenamefont {Niebergall},
  \citenamefont {Hashemi}, \citenamefont {Hergert},\ and\ \citenamefont
  {Stepanyuk}}]{hashemiphysica}%
  \BibitemOpen
  \bibfield  {author} {\bibinfo {author} {\bibfnamefont {P.~A.}\ \bibnamefont
  {Ignatiev}}, \bibinfo {author} {\bibfnamefont {N.~N.}\ \bibnamefont
  {Negulyaev}}, \bibinfo {author} {\bibfnamefont {L.}~\bibnamefont
  {Niebergall}}, \bibinfo {author} {\bibfnamefont {H.}~\bibnamefont {Hashemi}},
  \bibinfo {author} {\bibfnamefont {W.}~\bibnamefont {Hergert}}, \ and\
  \bibinfo {author} {\bibfnamefont {V.~S.}\ \bibnamefont {Stepanyuk}},\
  }\href@noop {} {\bibfield  {journal} {\bibinfo  {journal} {Phys. Status
  Solidi-B}\ }\textbf {\bibinfo {volume} {247}},\ \bibinfo {pages} {2537}
  (\bibinfo {year} {2010})}\BibitemShut {NoStop}%
\bibitem [{\citenamefont {Mokrousov}\ \emph
  {et~al.}(2007{\natexlab{b}})\citenamefont {Mokrousov}, \citenamefont
  {Bihlmayer}, \citenamefont {Bl\"{u}gel},\ and\ \citenamefont
  {Heinze}}]{mokrousov}%
  \BibitemOpen
  \bibfield  {author} {\bibinfo {author} {\bibfnamefont {Y.}~\bibnamefont
  {Mokrousov}}, \bibinfo {author} {\bibfnamefont {G.}~\bibnamefont
  {Bihlmayer}}, \bibinfo {author} {\bibfnamefont {S.}~\bibnamefont
  {Bl\"{u}gel}}, \ and\ \bibinfo {author} {\bibfnamefont {S.}~\bibnamefont
  {Heinze}},\ }\href {\doibase 10.1103/PhysRevB.75.104413} {\bibfield
  {journal} {\bibinfo  {journal} {Phys. Rev. B}\ }\textbf {\bibinfo {volume}
  {75}},\ \bibinfo {eid} {104413} (\bibinfo {year}
  {2007}{\natexlab{b}})}\BibitemShut {NoStop}%
\bibitem [{\citenamefont {B{\"u}rgi}\ \emph {et~al.}(2002)\citenamefont
  {B{\"u}rgi}, \citenamefont {Knorr}, \citenamefont {Brune}, \citenamefont
  {Schneider},\ and\ \citenamefont {Kern}}]{Blongrange}%
  \BibitemOpen
  \bibfield  {author} {\bibinfo {author} {\bibfnamefont {L.}~\bibnamefont
  {B{\"u}rgi}}, \bibinfo {author} {\bibfnamefont {N.}~\bibnamefont {Knorr}},
  \bibinfo {author} {\bibfnamefont {H.}~\bibnamefont {Brune}}, \bibinfo
  {author} {\bibfnamefont {M.}~\bibnamefont {Schneider}}, \ and\ \bibinfo
  {author} {\bibfnamefont {K.}~\bibnamefont {Kern}},\ }\href {\doibase
  10.1007/s003390101062} {\bibfield  {journal} {\bibinfo  {journal} {Applied
  Physics A}\ }\textbf {\bibinfo {volume} {75}},\ \bibinfo {pages} {141}
  (\bibinfo {year} {2002})}\BibitemShut {NoStop}%
\bibitem [{\citenamefont {Wahl}\ \emph {et~al.}(2007)\citenamefont {Wahl},
  \citenamefont {Simon}, \citenamefont {Diekh\"oner}, \citenamefont
  {Stepanyuk}, \citenamefont {Bruno}, \citenamefont {Schneider},\ and\
  \citenamefont {Kern}}]{PhysRevLett.98.056601}%
  \BibitemOpen
  \bibfield  {author} {\bibinfo {author} {\bibfnamefont {P.}~\bibnamefont
  {Wahl}}, \bibinfo {author} {\bibfnamefont {P.}~\bibnamefont {Simon}},
  \bibinfo {author} {\bibfnamefont {L.}~\bibnamefont {Diekh\"oner}}, \bibinfo
  {author} {\bibfnamefont {V.~S.}\ \bibnamefont {Stepanyuk}}, \bibinfo {author}
  {\bibfnamefont {P.}~\bibnamefont {Bruno}}, \bibinfo {author} {\bibfnamefont
  {M.~A.}\ \bibnamefont {Schneider}}, \ and\ \bibinfo {author} {\bibfnamefont
  {K.}~\bibnamefont {Kern}},\ }\href {\doibase 10.1103/PhysRevLett.98.056601}
  {\bibfield  {journal} {\bibinfo  {journal} {Phys. Rev. Lett.}\ }\textbf
  {\bibinfo {volume} {98}},\ \bibinfo {pages} {056601} (\bibinfo {year}
  {2007})}\BibitemShut {NoStop}%
\bibitem [{\citenamefont {Stepanyuk}\ \emph {et~al.}(2006)\citenamefont
  {Stepanyuk}, \citenamefont {Niebergall}, \citenamefont {Baranov},
  \citenamefont {Hergert},\ and\ \citenamefont {Bruno}}]{Stepanyuk2006272}%
  \BibitemOpen
  \bibfield  {author} {\bibinfo {author} {\bibfnamefont {V.}~\bibnamefont
  {Stepanyuk}}, \bibinfo {author} {\bibfnamefont {L.}~\bibnamefont
  {Niebergall}}, \bibinfo {author} {\bibfnamefont {A.}~\bibnamefont {Baranov}},
  \bibinfo {author} {\bibfnamefont {W.}~\bibnamefont {Hergert}}, \ and\
  \bibinfo {author} {\bibfnamefont {P.}~\bibnamefont {Bruno}},\ }\href
  {\doibase DOI: 10.1016/j.commatsci.2004.09.053} {\bibfield  {journal}
  {\bibinfo  {journal} {Comput. Mater. Sci.}\ }\textbf {\bibinfo {volume}
  {35}},\ \bibinfo {pages} {272 } (\bibinfo {year} {2006})},\ \bibinfo {note}
  {proceedings of the 4th International Conference on the Theory of Atomic and
  Molecular Clusters (TAMC-IV)}\BibitemShut {NoStop}%
\bibitem [{\citenamefont {Ding}\ \emph {et~al.}(2007)\citenamefont {Ding},
  \citenamefont {Stepanyuk}, \citenamefont {Ignatiev}, \citenamefont
  {Negulyaev}, \citenamefont {Niebergall}, \citenamefont {Wasniowska},
  \citenamefont {Gao}, \citenamefont {Bruno},\ and\ \citenamefont
  {Kirschner}}]{PhysRevB.76.033409}%
  \BibitemOpen
  \bibfield  {author} {\bibinfo {author} {\bibfnamefont {H.~F.}\ \bibnamefont
  {Ding}}, \bibinfo {author} {\bibfnamefont {V.~S.}\ \bibnamefont {Stepanyuk}},
  \bibinfo {author} {\bibfnamefont {P.~A.}\ \bibnamefont {Ignatiev}}, \bibinfo
  {author} {\bibfnamefont {N.~N.}\ \bibnamefont {Negulyaev}}, \bibinfo {author}
  {\bibfnamefont {L.}~\bibnamefont {Niebergall}}, \bibinfo {author}
  {\bibfnamefont {M.}~\bibnamefont {Wasniowska}}, \bibinfo {author}
  {\bibfnamefont {C.~L.}\ \bibnamefont {Gao}}, \bibinfo {author} {\bibfnamefont
  {P.}~\bibnamefont {Bruno}}, \ and\ \bibinfo {author} {\bibfnamefont
  {J.}~\bibnamefont {Kirschner}},\ }\href {\doibase 10.1103/PhysRevB.76.033409}
  {\bibfield  {journal} {\bibinfo  {journal} {Phys. Rev. B}\ }\textbf {\bibinfo
  {volume} {76}},\ \bibinfo {pages} {033409} (\bibinfo {year}
  {2007})}\BibitemShut {NoStop}%
\bibitem [{\citenamefont {Hashemi}(2015)}]{hashemithesis}%
  \BibitemOpen
  \bibfield  {author} {\bibinfo {author} {\bibfnamefont {H.}~\bibnamefont
  {Hashemi}},\ }\emph {\bibinfo {title} {First principles study of magnetic
  properties of nanowires on Cu surfaces}},\ \href@noop {} {Ph.D. thesis},\
  \bibinfo  {school} {Martin Luther University of Halle-Wittenberg} (\bibinfo
  {year} {2015})\BibitemShut {NoStop}%
\bibitem [{\citenamefont {Khajetoorians}\ \emph {et~al.}(2011)\citenamefont
  {Khajetoorians}, \citenamefont {Wiebe}, \citenamefont {Chilian},\ and\
  \citenamefont {Wiesendanger}}]{Khajetoorians1062}%
  \BibitemOpen
  \bibfield  {author} {\bibinfo {author} {\bibfnamefont {A.~A.}\ \bibnamefont
  {Khajetoorians}}, \bibinfo {author} {\bibfnamefont {J.}~\bibnamefont
  {Wiebe}}, \bibinfo {author} {\bibfnamefont {B.}~\bibnamefont {Chilian}}, \
  and\ \bibinfo {author} {\bibfnamefont {R.}~\bibnamefont {Wiesendanger}},\
  }\href {\doibase 10.1126/science.1201725} {\bibfield  {journal} {\bibinfo
  {journal} {Science}\ }\textbf {\bibinfo {volume} {332}},\ \bibinfo {pages}
  {1062} (\bibinfo {year} {2011})},\ \Eprint
  {http://arxiv.org/abs/http://science.sciencemag.org/content/332/6033/1062.full.pdf}
  {http://science.sciencemag.org/content/332/6033/1062.full.pdf} \BibitemShut
  {NoStop}%
\bibitem [{\citenamefont {Patrone}\ and\ \citenamefont
  {Einstein}(2012)}]{PhysRevB.85.045429}%
  \BibitemOpen
  \bibfield  {author} {\bibinfo {author} {\bibfnamefont {P.~N.}\ \bibnamefont
  {Patrone}}\ and\ \bibinfo {author} {\bibfnamefont {T.~L.}\ \bibnamefont
  {Einstein}},\ }\href {\doibase 10.1103/PhysRevB.85.045429} {\bibfield
  {journal} {\bibinfo  {journal} {Phys. Rev. B}\ }\textbf {\bibinfo {volume}
  {85}},\ \bibinfo {pages} {045429} (\bibinfo {year} {2012})}\BibitemShut
  {NoStop}%
\bibitem [{\citenamefont {Simon}\ \emph {et~al.}(2011)\citenamefont {Simon},
  \citenamefont {\'Ujfalussy}, \citenamefont {Lazarovits}, \citenamefont
  {Szilva}, \citenamefont {Szunyogh},\ and\ \citenamefont
  {Stocks}}]{PhysRevB.83.224416}%
  \BibitemOpen
  \bibfield  {author} {\bibinfo {author} {\bibfnamefont {E.}~\bibnamefont
  {Simon}}, \bibinfo {author} {\bibfnamefont {B.}~\bibnamefont {\'Ujfalussy}},
  \bibinfo {author} {\bibfnamefont {B.}~\bibnamefont {Lazarovits}}, \bibinfo
  {author} {\bibfnamefont {A.}~\bibnamefont {Szilva}}, \bibinfo {author}
  {\bibfnamefont {L.}~\bibnamefont {Szunyogh}}, \ and\ \bibinfo {author}
  {\bibfnamefont {G.~M.}\ \bibnamefont {Stocks}},\ }\href {\doibase
  10.1103/PhysRevB.83.224416} {\bibfield  {journal} {\bibinfo  {journal} {Phys.
  Rev. B}\ }\textbf {\bibinfo {volume} {83}},\ \bibinfo {pages} {224416}
  (\bibinfo {year} {2011})}\BibitemShut {NoStop}%
\bibitem [{\citenamefont {Brovko}\ \emph {et~al.}(2008)\citenamefont {Brovko},
  \citenamefont {Stepanyuk},\ and\ \citenamefont {Bruno}}]{PhysRevB.78.165413}%
  \BibitemOpen
  \bibfield  {author} {\bibinfo {author} {\bibfnamefont {O.~O.}\ \bibnamefont
  {Brovko}}, \bibinfo {author} {\bibfnamefont {V.~S.}\ \bibnamefont
  {Stepanyuk}}, \ and\ \bibinfo {author} {\bibfnamefont {P.}~\bibnamefont
  {Bruno}},\ }\href {\doibase 10.1103/PhysRevB.78.165413} {\bibfield  {journal}
  {\bibinfo  {journal} {Phys. Rev. B}\ }\textbf {\bibinfo {volume} {78}},\
  \bibinfo {pages} {165413} (\bibinfo {year} {2008})}\BibitemShut {NoStop}%
\bibitem [{\citenamefont {Stepanyuk}\ \emph {et~al.}(2003)\citenamefont
  {Stepanyuk}, \citenamefont {Baranov}, \citenamefont {Tsivlin}, \citenamefont
  {Hergert}, \citenamefont {Bruno}, \citenamefont {Knorr}, \citenamefont
  {Schneider},\ and\ \citenamefont {Kern}}]{PhysRevB.68.205410}%
  \BibitemOpen
  \bibfield  {author} {\bibinfo {author} {\bibfnamefont {V.~S.}\ \bibnamefont
  {Stepanyuk}}, \bibinfo {author} {\bibfnamefont {A.~N.}\ \bibnamefont
  {Baranov}}, \bibinfo {author} {\bibfnamefont {D.~V.}\ \bibnamefont
  {Tsivlin}}, \bibinfo {author} {\bibfnamefont {W.}~\bibnamefont {Hergert}},
  \bibinfo {author} {\bibfnamefont {P.}~\bibnamefont {Bruno}}, \bibinfo
  {author} {\bibfnamefont {N.}~\bibnamefont {Knorr}}, \bibinfo {author}
  {\bibfnamefont {M.~A.}\ \bibnamefont {Schneider}}, \ and\ \bibinfo {author}
  {\bibfnamefont {K.}~\bibnamefont {Kern}},\ }\href {\doibase
  10.1103/PhysRevB.68.205410} {\bibfield  {journal} {\bibinfo  {journal} {Phys.
  Rev. B}\ }\textbf {\bibinfo {volume} {68}},\ \bibinfo {pages} {205410}
  (\bibinfo {year} {2003})}\BibitemShut {NoStop}%
\bibitem [{\citenamefont {Stepanyuk}\ \emph {et~al.}(2004)\citenamefont
  {Stepanyuk}, \citenamefont {Niebergall}, \citenamefont {Longo}, \citenamefont
  {Hergert},\ and\ \citenamefont {Bruno}}]{PhysRevB.70.075414}%
  \BibitemOpen
  \bibfield  {author} {\bibinfo {author} {\bibfnamefont {V.~S.}\ \bibnamefont
  {Stepanyuk}}, \bibinfo {author} {\bibfnamefont {L.}~\bibnamefont
  {Niebergall}}, \bibinfo {author} {\bibfnamefont {R.~C.}\ \bibnamefont
  {Longo}}, \bibinfo {author} {\bibfnamefont {W.}~\bibnamefont {Hergert}}, \
  and\ \bibinfo {author} {\bibfnamefont {P.}~\bibnamefont {Bruno}},\ }\href
  {\doibase 10.1103/PhysRevB.70.075414} {\bibfield  {journal} {\bibinfo
  {journal} {Phys. Rev. B}\ }\textbf {\bibinfo {volume} {70}},\ \bibinfo
  {pages} {075414} (\bibinfo {year} {2004})}\BibitemShut {NoStop}%
\bibitem [{\citenamefont {Lau}\ and\ \citenamefont {Kohn}(1978)}]{LAU197869}%
  \BibitemOpen
  \bibfield  {author} {\bibinfo {author} {\bibfnamefont {K.}~\bibnamefont
  {Lau}}\ and\ \bibinfo {author} {\bibfnamefont {W.}~\bibnamefont {Kohn}},\
  }\href {\doibase https://doi.org/10.1016/0039-6028(78)90053-5} {\bibfield
  {journal} {\bibinfo  {journal} {Surface Science}\ }\textbf {\bibinfo {volume}
  {75}},\ \bibinfo {pages} {69 } (\bibinfo {year} {1978})}\BibitemShut
  {NoStop}%
\bibitem [{\citenamefont {Ignatiev}\ \emph {et~al.}(2007)\citenamefont
  {Ignatiev}, \citenamefont {Stepanyuk}, \citenamefont {Klavsyuk},
  \citenamefont {Hergert},\ and\ \citenamefont {Bruno}}]{PhysRevB.75.155428}%
  \BibitemOpen
  \bibfield  {author} {\bibinfo {author} {\bibfnamefont {P.~A.}\ \bibnamefont
  {Ignatiev}}, \bibinfo {author} {\bibfnamefont {V.~S.}\ \bibnamefont
  {Stepanyuk}}, \bibinfo {author} {\bibfnamefont {A.~L.}\ \bibnamefont
  {Klavsyuk}}, \bibinfo {author} {\bibfnamefont {W.}~\bibnamefont {Hergert}}, \
  and\ \bibinfo {author} {\bibfnamefont {P.}~\bibnamefont {Bruno}},\ }\href
  {\doibase 10.1103/PhysRevB.75.155428} {\bibfield  {journal} {\bibinfo
  {journal} {Phys. Rev. B}\ }\textbf {\bibinfo {volume} {75}},\ \bibinfo
  {pages} {155428} (\bibinfo {year} {2007})}\BibitemShut {NoStop}%
\bibitem [{\citenamefont {Mermin}\ and\ \citenamefont
  {Wagner}(1966)}]{PhysRevLett.17.1133}%
  \BibitemOpen
  \bibfield  {author} {\bibinfo {author} {\bibfnamefont {N.~D.}\ \bibnamefont
  {Mermin}}\ and\ \bibinfo {author} {\bibfnamefont {H.}~\bibnamefont
  {Wagner}},\ }\href {\doibase 10.1103/PhysRevLett.17.1133} {\bibfield
  {journal} {\bibinfo  {journal} {Phys. Rev. Lett.}\ }\textbf {\bibinfo
  {volume} {17}},\ \bibinfo {pages} {1133} (\bibinfo {year}
  {1966})}\BibitemShut {NoStop}%
\end{thebibliography}%

\end{document}